# Applications of Data Mining Techniques for Vehicular Ad hoc Networks

Mohammed AL Zamil and Samer Samarah
Department of Computer Information Systems
Yarmouk University
Mohammedz@yu.edu.jo, Samers@yu.edu.jo

## Abstract

Due to the recent advances in vehicular ad hoc networks (VANETs), smart applications have been incorporating the data generated from these networks to provide quality of life services. In this paper, we have proposed taxonomy of data mining techniques that have been applied in this domain in addition to a classification of these techniques. Our contribution is to highlight the research methodologies in the literature and allow for comparing among them using different characteristics. The proposed taxonomy covers elementary data mining techniques such as: preprocessing, outlier detection, clustering, and classification of data. In addition, it covers centralized, distributed, offline, and online techniques from the literature.

# 1. Introduction

Recent advances in hardware technologies and wireless communication have led to the development of powerful generations of sensors that is able to read and transport environmental variables and observations. However, such generations play a significant role in designing smart applications, which, in turns, serves the development of high quality and safety services. Because of its pervasive surveillance capabilities, wireless sensor networks have been implemented in many application domains such as Medical Assistant Systems [1, 2, 3, 4, 5]. Recently, vehicular ad hoc networks have been introduced and implemented as a special type of wireless sensor networks to provide smart applications for managing traffic in modern societies.

Vehicular Ad-hoc Network (VANET) is a smart technology that integrates the capabilities of mobile ad hoc networks (MANET) with smart vehicles [6, 7, 8]. VANETs are designed to provide means of communication among vehicles, where communication entities (vehicles) changing their location continuously. Thus, VANETs raise the spatiotemporal issues that directly affect its performance in terms of accuracy and quality of life services in traffic managements systems.

VANETs systems process a huge (non-decreasing) amount of changing, geographically dispersed, homogeneous and heterogeneous data. Analyzing such elementary information will facilitate automatic prediction, control, and decision making. Therefore, mining VANETs raw data for recognizing patterns of action, outliers, clusters, and classifications is an emergent research application to develop intelligent traffic management systems.

Recent advances in data mining have considered the tremendous amount of that have been generated by vehicular networks. Previous researches in data mining have proposed many techniques, which suggested different data analysis tasks to handle the explosion of data generated by VANETs [9, 10, 11]. Although VANETs are special type of wireless sensor networks, the ad hoc nature of distributed sensors and the spatiotemporal aspects of its constituent nodes make it difficult to adapt data analysis solutions from wireless sensor networks. In addition, modern vehicular networks force emergent challenges as compared to existing sensory networks. These challenges are summarized into two aspects: 1) VANETs are not affected by energy consumption, 2) sensors are moving over time; not stable.

However, the architectural setup of VANET generates new research challenges because of its dynamicity, random distribution of sensors, and the effect of the environmental variables on the accuracy of the resulted analysis. For instance, road specifications have a direct effect on concluding behavioral patterns. These challenges make traditional mining techniques, even these techniques for wireless sensor networks, inapplicable because traditional techniques are in appropriate in terms of accuracy, performance, and communication overhead due to the dynamic and spatiotemporal nature of VANETs environments. Therefore, many techniques have been proposed to handle the unique aspects of VANETs

For example, outlier detection, clustering, and classification techniques for wireless sensor networks that have been introduced in [12, 13, 14, 15, 16, 17] are not applicable in a dynamic and ad hoc networks such as VANETs as these techniques assumed that wireless sensors are non-moving nodes. For instance, VANETs nodes may change their cluster from time to time as nodes moving from one location to another.

Several surveys and reviews have been examined data mining techniques on different domains of knowledge such as medical domain [18], data stream analysis for extracting frequent patterns [19, 20], clustering algorithms for wireless sensor networks [21, 22], data stream classification techniques [23], improving software interfaces [24], design and searching for patterns [25, 26, 27], verification of sensory networks [28, 29, 30], human activity recognition [31, 32], energy consumption [33], and data mining techniques for wireless sensor networks [34]. However, none of the above reviews and surveys examined data mining techniques that concentrate on a taxonomy for mining techniques applied on Vehicular Ad hoc Networks systems.

This review paper introduces taxonomy of data mining techniques that can be applied for VANETs data mining and analysis. The paper summarizes state-of-the-art algorithms and techniques that have been designed for vehicular ad hoc networks and similar technologies. Furthermore, an evaluation, applicability, and limitations of every technique are presented. Finally, the paper briefly describes plenty of research problems that could be solved by applying data mining techniques for VANETs.

The rest of the paper is organized as follows: section 2 highlights the challenges of traditional data mining tasks on data generated from wireless networks. Section 3 presents the proposed taxonomy of existing data mining techniques with respect to its applicability on wireless networks. Section 4 provides an analysis of existing methodologies against the proposed taxonomy. Section 5 compares among existing data mining techniques on WSN in terms of different aspect. Section 6 illustrates the limitations of the proposed taxonomy. Finally, sections 7 and 8 discuss the future research and conclude the importance of the proposed taxonomy.

## 2. Background Knowledge

### 2.1 Data Mining Tasks

Data mining is defined as the process of analyzing large amount of data for the purpose of extracting new knowledge [35]. In general, there are seven essential steps to perform knowledge discovery using data mining techniques. Figure 1 shows these tasks. Data cleaning is the process of dealing with noisy and inconsistent data by removing or reproducing dirty ones. Once the data is clean, integration becomes an essential task in situations where multiple sources of data are participating to form huge repository. However, huge repositories might contain large amount of attributes. These attributes represent the properties of entities as the basic constituent to the data source. In data mining, it is necessary to select some attributes, according to the application requirement, for the analysis process. For instance, information-gain theory is one of the common algorithms to select representative features.

In order to make the selected data in an appropriate form that is accepted by some mining techniques, data transformation could be applied such as summarization or aggregation. Next, intelligent algorithms are applied for the purpose of extracting patterns of data. The sixth task focusing on evaluating resulted patterns using usefulness measurements. Such evaluation represents the performance of a data mining technique. Finally, knowledge presentation techniques are applied in order to visualize or represent knowledge to intended users.

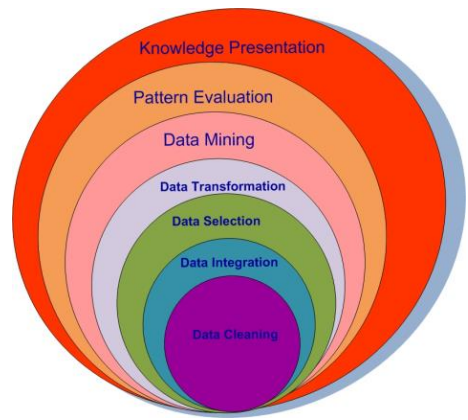

Figure 1 Common Data Mining Tasks

## 2.2 Traditional Vs Domain-Specific Tasks for VANETs

Traditional data mining techniques have built to serve centralized processing, in which data is processed at a central unit and, then, the conclusions are distributed to connecting nodes. This architecture allows fast processing as, usually; the central unit has powerful computational power. In addition, these techniques rely on constructing models based on static (non-changing) features. Recent research on big data analysis [36] showed that traditional techniques requires more and more computational time, which could make it useless in smart and modern applications.

Specialized domain data mining techniques for VANETs have focused on models patterns of data to serve specific application with acceptable performance from non-decreasing data streams from VANETs nodes. Such techniques should consider the nature of collected data; including:

- Spatiotemporal aspects : nodes are changing their location continuously
- Ad hoc nature: nodes movements are not predictable
- Distribution: distributed solutions enhance computational time and storage requirements
- Dynamicity: features are not static; data values are dynamic
- High Response time: un timely decisions lose their value

Traditional data mining techniques have not designed to handle such issues for mining and analyzing data. Therefore, there is a need for specialized techniques that are able to provide smart and intelligent solutions for mining and analyzing data collected from vehicular ad hoc networks. Table 1 summarizes the differences between traditional and VANETs data mining techniques.

Table 1 Differences between Traditional and Specialized Techniques for VANETs

| Characteristic | Traditional Techniques | VANET-Specialized Techniques |
|---|---|---|
| Centralization | Centralized | Distributed |
| Nature of Data | Static | Dynamic |
| Features | Fixed | Changing |
| Node Location | Fixed | Dynamic |
| Response Time | Asynchronous | Synchronous |
| Time Complexity | High | Low |

## 2.3 Research Challenges

Section 2.2 lists the differences between the characteristics of traditional data mining techniques and specialized techniques for VANETs. The comparison has been made based on the required aspects of domain applications. In this section, we illuminate the new research challenges resulted from such differences.

1. High performance and accurate algorithms are required while preserving or minimizing the required computation. VANETs applications require timely and accurate information as late information loss their value (Time Complexity).
2. Fully or semi Distribution of data processing so that nodes might take the decision by their own. Pure centralized data processing is useless for reasons such as: bottle-nick, single-point-of-failure, and performance degradation.
3. Algorithms should integrate static and dynamic data; static data to preserve the environmental characteristics (such as roads) and dynamic data to deal with node movements.
4. The data mining feature extraction task should take into account that some features might leave and some might come as objects are moving over time from environment to another.
5. The response time challenge requires developing online mining and analysis of data; such as: online outlier detection and online clustering and classification of moving data objects.
6. Because environmental variables and mining features are subject to change, the mining results might change from time to time. This implies that the mining model should adaptable so that it could be changed frequently. Capturing the changing results over time is a key challenge in this domain.
7. As the number of nodes in VANETs is not predictable (ad hoc) and they might be subject to similar circumstances, data aggregation and transformation techniques play a significant role for reducing network traffics.

8. In VANETs, the nature of data is not fixed; the components o f the network might generate homogenous and heterogeneous data. Integrating, analyzing, and mining patterns from both source of dynamic data is a key challenge.

During the last decade, researchers have developed new techniques by modifying the traditional ones to cope with VANETs data characteristics. In this paper, we organized these techniques according to a proposed taxonomy that classify the data mining techniques for VANETs into set of categories based on the data mining task.

## 3. Taxonomy of Surveyed Techniques

This section introduces our taxonomy for current techniques that have been proposed for mining data generated by VANETs. Our approach relies on top-down taxonomy, in which abstract levels are representing general and state-of-the-art techniques. However, lower levels are dedicated for vehicular networks to handle its own aspects and specifications such as its dynamicity and topology stability. Figure 2 shows our proposed taxonomy.

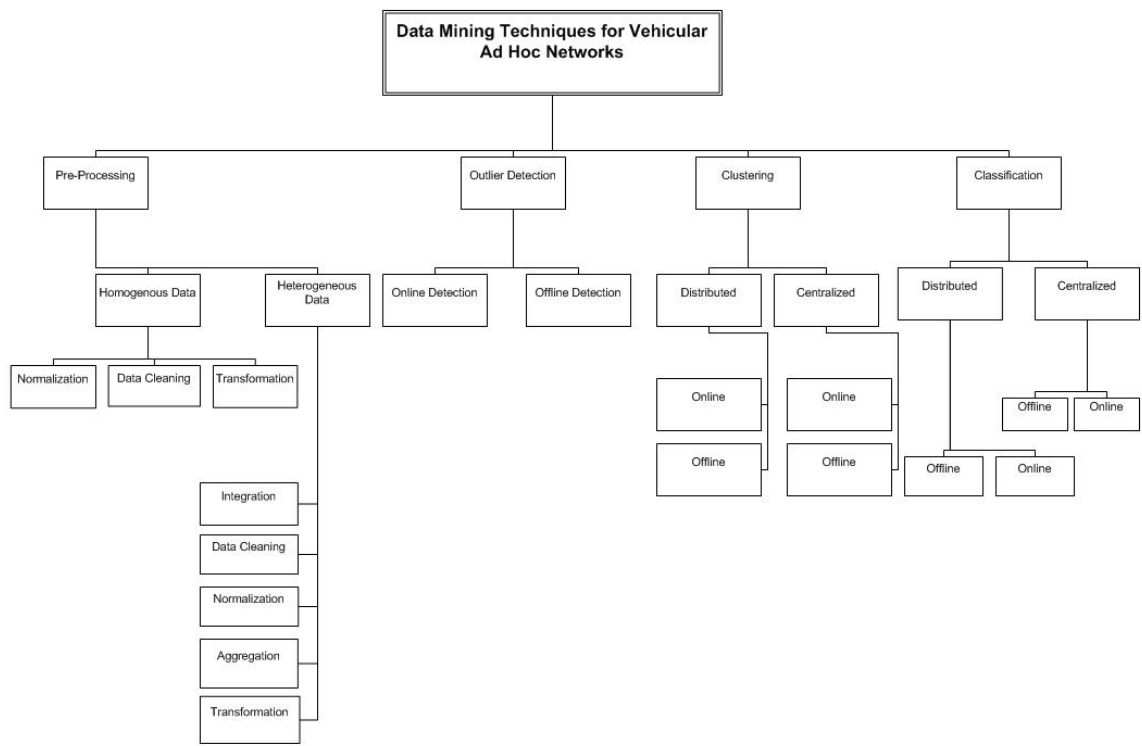

Figure 2 Taxonomy of Data Mining Techniques for VANETs

Our taxonomy refers to four data mining tasks: preprocessing, outlier detection, clustering, and classification techniques. First, we considered the preprocessing task since it is highly significant especially when the sources of data are heterogeneous. In this case, data need to be integrated, cleaned, normalized, aggregated (in some cases), and transformed into a unified form. On the other hand, homogenous sources also need to be cleaned, normalized and transformed in order to come up with unified services.

Next, outlier detection plays a significant role during the development of VANETs services. It is applied to filter out intruders and separate operational regions. Two types of outlier detection are investigated: online and offline techniques. While online techniques are prone to errors and performance degradation, they are important in alarming emergent situations. On the other hand, offline outlier detection techniques are important to analyze histories and report intrusions or patterns of penetrations.

Clustering of vehicles is a key service during the system lifetime. Within the same road, vehicles are subject to be categorized for customized services or creating an instance topology. In this taxonomy, we found that clustering techniques are of two types: centralized with a cluster head or dynamically distribute the decision power to all vehicles (nodes). Both techniques are well-known in the computer science. However, customizing such techniques in VANETs requires careful design of online and offline techniques.

Finally, predicting a class label for vehicles is a key service. It is the process of assigning a label to one or group of vehicles. There many classification techniques that have been proposed to handle such task. Most of these techniques are centralized due to the difficulty to distribute historical data, which is required to supervise the learning process. Furthermore, online classification remains a challenge in this domain, with noticeable difficulty in achieving acceptable performance.

## 4. Data Mining Techniques for VANETs

### 4.1 Data Pre-Processing

In this section, a set of data mining techniques are illuminated to identify how preprocessing tasks have solved several problems in VANETs. However, content sharing schemes have been studied [37] to overcome unreliable connections, spatiotemporal aspects, and interoperability among vehicles in VANETs.

This section investigates different preprocessing techniques that have developed for cleaning, transforming, or normalizing data collected from VANETs. Because VANETs might run among different manufacturing vehicles with different standards and platforms, researchers differentiate among homogenous and heterogeneous preprocessing tasks. This taxonomy makes it easier to understand the complexity of applying techniques.

In [38], authors proposed a novel technique, called RD: Role-Differentiated, to detect false data in VANETs. This technique assumes that every sensor is able to attach a confidence score for every reporting event; this depends on rules defined by the standard operator of the VANETs (i.e. traffic department). According to the confidence scores reported by sensors in a VANET, the proposed algorithm computes the plausibility of a given event based on an empirical mathematical model.

Detecting of missing values is a key task since wireless communications are prone to noise. Many solutions have been proposed such as: predicting missing values using statistical dispersion measure [39], applying association rules induced from known patterns based on closed-frequent item sets [40], and applying precision estimations using physical models (DEPM) or statistical models (DESM) to examine specification of sensed features [41].

Another interested technique for handling noisy data is the one proposed by [42], which handled the estimation of data according to the spatial characteristics of planted sensors using graph mappings. Rather, researchers in [43] proposed a methodology that is based on the connectivity among sensors to develop association rules for predicting missing values.

In [44] researchers applied multiple-regression model for predicting missing values. First, they applied linear regression model to predict spatiotemporal dimensions. Then, they assign the weighted coefficients to the two spatiotemporal estimated values computed according to the goodness-of-fit, and then uses the weighted average of the two values as the final predicted value

In [45], researchers studied traffic information queries in VANETs. They proposed an efficient querying algorithm with restricted conditions to broadcast querying information to vehicles around. Their algorithm reduces the redundant data and control the communication congestion effectively.

In [46], researchers proposed a method for building dynamic data for solving data aggregation problem and minimizing the number data sources. In [47], researchers proposed Caas to structure the network into a set of clusters according to some clustering features. The purpose is to provide content based routing for communication among clusters and delay routing for communication among external clusters.

In [48], researchers provided an extensive study to the authentication problem in VANETs, which is required to validation vehicle-to-vehicle, vehicle-to-infrastructure, and infrastructure-to-vehicle communications. The paper show different authentication algorithms and cryptographic schemes.

In [49], researchers studied data aggregation in wireless sensor networks to reduce redundant data and improve communication. The paper introduced an adaptive forwarding delay control scheme, namely Catch-Up, which dynamically changes the forwarding speed of nearby reports so that they have a better chance to meet each other and be aggregated together. In [50], researchers proposed MCNC (Multicast Network Coding) for aggregating and disseminating data in VANETs.

In [51], researchers proposed an aggregation model implemented in road side units using time converge globally and space converge locally. This method reduces data redundancy and transmission delay. In [52], researchers studied the security models for a set data aggregation models in VANETs.

## 4.2 Outlier Detection

Work on outlier detection including misbehavior, intruder, and anomaly detection has produced very rich literature in VANETs and mobile ad hoc networks. This section summarizes and discusses the state-of-the-art techniques developed for VANETs. To facilitate understanding the outlier problem in VANETs, the reviewed techniques have been divided into offline and online detection techniques. The former represents these techniques that build to respond in an asynchronous manner, while the later respond on time (synchronous).

In the literature, the outlier detection techniques have been classified into four categories [53]: failed node behavior, badly failed node behavior, selfish attack, and malicious attack. The previous classification based on the node's intent and action during communication. Selfish attack is defined as intentional passive misbehavior in which nodes do not fully participate in forwarding messages to conserve their resources. On the other hand, malicious attack is defined as an intentional active misbehavior in which nodes aims to interrupt network operations.

Many intrusion detection systems (IDS) have been built to handle different node misbehaviors. Several techniques have been proposed to develop such systems on individual peers due to the lack of standardize infrastructure [54, 55, 56, 57]. Most of these techniques rely on embedding nodes with IDS sensor that is considered always an active sensor. Such assumption raises the problem of energy consumption and its important effect on various wireless sensor networks technologies. However, research in Huang et al. [58] propose a an IDS framework, in which detecting misbehavior nodes according to their clusters that are formed in ad hoc networks, which lead to reduce the power consumption for each node.

On the common malicious activities in ad hoc networks is routing misbehavior, where attacker aims to pollute the ad hoc network in order to plant some nodes. Then, the attacker uses these nodes to disturb the routing services. An interesting research to handle this kind of attack has been introduced by Marti et al. [59], which proposed two techniques, namely *watchdog and pathrater* , to detect and isolate misbehaving nodes, which are nodes that do not forward packets. Furthermore, many other solutions have been proposed to handle routing attacks such as [60, 61, 62].

Raya et al. [63] proposed a trust management scheme for VANET, in which a distributed spanning tree (DST) is used to combine multiple evidences for trust. In their paper, DST is not applied to combine evidences in real-time. On the other hand, DST has been used to integrate the direct observations from each IDS sensor in [64]. In [65], researchers proposed a weighted voting technique for detecting misbehavior nodes.

In [66] researchers proposed a multi-model technique that provide the ability to detect outliers and model inconsistencies. In [67], researchers proposed RaBTM; a beacon-based trust management system that prorogate opinions and block internal attackers from sending or forwarding forged packets in VANETs. A similar but self-organized trust management system for VANETs has been proposed in [68].

## 4.3 Data Clustering

Data clustering is defined as the process of categorizing related data into a set of clusters. It is an unsupervised learning; in which the clustering algorithm has no clue about the relationships among data objects. Two directions have been considered: data clustering and node clustering [69]. There many proposed approaches in the literature that handle the problem using data or node clustering, each of them has its advantages and disadvantages. In this section, we highlight these techniques according to our proposed taxonomy.

H-Cluster has been proposed by [70], which proposes a node based clustering by promoting the idea of local and global clusters. The local clusters combine sensors in which their features can be combined. A similar algorithm (Energy efficient dynamic clustering EEDC) proposed in [71] that focusing on clustering sensor nodes, but for the purpose of minimizing power consumption. EEDC is functioning on grouping homogeneous sensors in order to find alternatives.

LEACH [72], HEED [73], and MRECA[74] are well-known techniques in clustering sensor nodes for the purpose of minimizing the power consumption of the sensory network. While these algorithms are focusing on scalability as well as energy, all of them require dynamic topology schemes when applying to VANETs. They all designed to organize unmoving sensors, rather than moving ones.

Another interested approach that has been proposed in [75] relies on measuring the distance between the cluster head and its candidate members. It is relying on the spatio-temporal features of the leader node by collecting such features into vectors and then measures the distance or the statistical correlation among them. The nearest group is the highly candidate one. The only limitation of this algorithm is the characteristics of the cluster head and the way the algorithm voting for it. In VANETs, the dynamicity enforces changing the cluster head frequently. DCC (data correlation-based clustering) [76] is a similar approach, but based on data clustering; rather than nodes.

Data-based clustering techniques were relying on aggregation [77], and queries generated from nodes (similarity) [78]. The performance of applying such techniques on VANETs data are not acceptable since all these techniques missed the fact that objects are moving overtime; thus the space and time features should be incorporated.

A distributed, hierarchical clustering and Summarization algorithm (DHCS) [79] has been proposed for clustering online data generated from cordless devices such as VANETs. The proposed technique clusters vehicles by combining data with their geographical approximate.

## 4.4 Data Classification

Data classification is a supervised learning task in data mining in which a predefined set of data (training) is used to build a model then tested (testing data) against classification accuracy. It is supervised since the training and testing data sets are labeled, and assumed to be correct. Many applications can be extracted from classifying VANET data, such as: detecting emergent situations, road status, and predicting future events.

Rule-based classification techniques are the simplest classifiers that are easily to modeled and apply. In [23] a simple lazy classifier has been proposed in a framework that can be applied to VANET data. Another approach to facilitate the generation of rules is the application of decision trees. In [23], researchers proposed the development of a decision tree using frequent-pattern generation. On the other hand, a similar approach in [19] applies Apriori algorithm to extract the frequent-patterns.

Trajectory classification shows a significant performance in classifying moving objects. In [80], authors proposed the application of nearest neighbor trajectory classification (NNTC) using frequent detected labels. A major limitation of this technique is the ignorance of the environmental variables. Such limitation has been handled by [81] using a prediction model that incorporate environmental variables that are subject to change overtime. However, in many cases, data granularity plays a significant role in degrading the performance of the classifier. Researchers in [82] proposed lightweight technique to minimize the effect of granularity using one-pass algorithm.

Implementing classifiers in a distributed environment has also attracted researchers in this domain. A voting-based distributed classifier has been proposed in [83] for the purpose of incorporating and aggregate information from all sensors in the network. While this approach seems heavy and hard to implement in a changing environment, it provide high performance in terms of accuracy and detection of emergent situations.

In [84], authors suggest the segmentation of vectors according to fixed and weighted segments. The proposed incremental learning process contributes in enhancing the performance while minimizing error rates. The limitation of this work is the implementation complexity, since the algorithm spans all possible combination of vectored features in a sequential order.

# 5. Comparison of Some Existing Clustering and Classification Techniques

In this section, we provide a comparison among different clustering and classification techniques. These comparisons have been conducted on data generated by VANET simulator (NS2). Table 2 shows different algorithms and their performance in terms of the average accuracy, intra-cluster balancing, and time complexity

Table 2 A comparison among different clustering techniques against accuracy and time complexity

| Cluster Algorithm | Average of Correctly Clustered Instances | Average of Within-Cluster Sum of Squared Error | Time Complexity |
|---|---|---|---|
| HEED | 54.9% | 21.4% | $O(N^2)$ |
| MRECA | 49.1% | 28.6% | $O(N)$ |
| EEDC | 56.4% | 19.5% | $O(N.log_n.K)$ |
| LEACH | 63.2% | 15.7% | $O(N.log_n.K)$ |
| Dynamic Rough-based Cluster | 68.1% | 11.6% | $O(N.K)$ |

Table 3 shows state-of-the-art classification techniques that have been widely applied to VANET datasets. The comparison is based on reporting the accuracy measurements of TP, FP, and TN.

Table 3 A comparison of state-of-the-art classifiers against accuracy

| Classifier | True-Positive | False-Positive | True-Negative |
|---|---|---|---|
| C 4.5 (J48 based) | 68.3% | 28.6% | 21.4% |
| SVM (Support Vector Machine) | 68.6% | 28.2% | 20.8% |
| K-NN (K-Nearest Neighbor) | 60.7% | 31.4% | 22.2% |
| MLP (A Multilayer Perceptron) | 69.1% | 26.3% | 21.9% |
| LPM (Linear Programming Machine) | 59.2% | 38.6% | 28.1% |
| RDA (Regularized Discriminant Analysis) | 57.5% | 39.9% | 29.8% |
| FD (Feature-Deselective) | 51.4% | 42.0% | 33.5% |
| DEC | 79.9% | 20.1% | 10.0% |
| (Dynamic Event) | | | |

# 6. Conclusion

This paper provided a survey on key techniques and algorithms that have been proposed in the literature and covered the data mining techniques that have been applied to wireless sensor networks and vehicular ad hoc networks. The contribution of this paper is to propose a taxonomy that is able to classify such techniques according to their applicability and requirements.

It is also provides researchers with brief information and literature that would help in advancing the research in analyzing VANET data. As a result, a comparison of different algorithmic characteristics has been provided, which would facilitate choosing algorithms according to the problems under investigation.